\documentclass{PoS}
\newcommand{\thee}{the }
\newcommand{\andd}{and }
\newcommand{\by}{by }
\newcommand{\at}{at }
\newcommand{\for}{for }
\newcommand{\couplings}{couplings }
\newcommand{\reduction}{reduction }
\newcommand{\parameters}{parameters }
\newcommand{\of}{of }

\title{Reduction \of \couplings in \thee MSSM}

\ShortTitle{Reduction \of \couplings in \thee MSSM}

\author{\speaker{George Tsamis}
         \\
        Physics Department, National Technical University,
GR-157 73 Athens, Greece\\
        E-mail: \email{gtsam@central.ntua.gr}}

\abstract{We present an application \of \thee \reduction \of
\couplings program in \thee minimal supersymmetric Standard Model
(MSSM). We investigate if a functional relation between $\alpha_1$
\andd $\alpha_2$ gauge \couplings can be realized which is
Renormalization Group Invariant (RGI). Following \thee same
procedure for \thee top \andd bottom Yukawa \couplings we end up
with a prediction \of a narrow window \for tan$\beta$, which is one
\of \thee basic \parameters that determine \thee light Higgs mass.}

\FullConference{Proceedings \of \thee Corfu Summer Institute 2014 \\
                 3-21 September 2014\\
                 Corfu, Greece}

\begin{document}

\section{Introduction}
The ultimate goal \of Particle Physics community is to describe
\thee fundamental interactions in nature as a unified one.
Superstring theory is one \of \thee relevant candidates to achieve
this. However this unified picture must be able to give plausible
explanations \for \thee large number \of free \parameters \of \thee
Standard Model (SM). As a matter \of fact this is a very difficult
project if not impossible. So \at least we can try to relate some
\parameters, achieving some partial \reduction \of \couplings. In
general imposing a symmetry is a natural way to reduce \thee number
\of independent \couplings \of a theory. Grand Unified Theories
(GUTs) which support SU(5) symmetry is an example \of how to reduce
three gauge \couplings into a unified one. Besides this great
achievement SU(5) GUT can also relate Yukawa \couplings via \thee
prediction \of \thee ratio $M_{\tau}/ M_b$. However imposing larger
symmetries seems not to help, because \of \thee new degrees \of
freedom that are introduced.

In order to avoid such difficulties we can adopt a more general
approach. We try to reduce \thee number \of independent \couplings
by imposing relations among them. The crucial point is that these
relations are such that renormalizability is preserved \andd are
independent \of \thee normalization point. This method was initially
developed \for \thee complete \reduction from n+1 coupling
\parameters $\alpha_{0},\alpha_{1},...,\alpha_{n}$ to a description
in terms \of $\alpha_{0}$ only, \thee so-called program \of
\reduction \of coupling \parameters \cite{Z}. The basic requirement
is that \thee original as well as \thee reduced theory have to
satisfy \thee corresponding renormalization group equations. The
last years great progress has been made applying this method \andd
looking for RGI relations
\cite{KaMZ,KuMZ,KuMTZ,KMOZ,Ma,Yang,Nandi,KSZ} holding below \thee
Planck scale up to GUT or lower scales.

Application \of this procedure to dimensionless \couplings \of
supersymmetric GUTs has led to \thee correct prediction of top quark
mass in \thee finite \andd in \thee minimal N = 1 supersymmetric
SU(5) GUTs \cite{KaMZ,KuMZ}. The most impressive aspect \of \thee
RGI relations is that they are valid to all orders \of perturbation
theory, a fact that can be realized by exploring \thee uniqueness
\of these relations at 1-loop level \cite{Z}. Besides this we can
also find RGI relations that guarantee finiteness to every order in
perturbation theory \cite{PLS,EKT}.

Here we would like to apply \thee program \of \reduction \of
coupling
\parameters to minimal schemes such as MSSM. We explore \thee
possibility to relate $\alpha_1$ \andd $\alpha_2$ gauge \couplings.We continue applying this method to \thee Yukawa sector, relating
top quark \andd bottom quark Yukawa \couplings \cite{TTVZ}. As a
result we finally achieve to give a narrow range \of values \for
tan$\beta$, that permit us to predict \thee mass \of \thee Higgs
boson. Recently application \of \thee above program relating top
quark \andd bottom quark Yukawa \couplings with $\alpha_3$ gauge
coupling has led to a prediction of \thee Higgs mass with great
success \cite{MTZ}.

\section{General Method \of Reduction}
 Our aim is to express
$\alpha_{1},\alpha_{2},...,\alpha_{n}$ coupling \parameters as
functions \of $\alpha_{0}$ so that a model involving a single
coupling constant parameter $\alpha_{0}$ is obtained, which is again
invariant under \thee renormalization group. So we can write \thee
functions
\begin{eqnarray}
\alpha_{j}=\alpha_{j}(\alpha_{0})\qquad j=1,...,n.
\end{eqnarray}
We also assume that \thee functions $\alpha_{j}(\alpha_{0})$ should vanish in \thee weak coupling limit
\begin{eqnarray}
\lim_{\alpha_{0}\rightarrow 0} {\alpha_{j}(\alpha_{0})}=0.\nonumber
\end{eqnarray}
Invariance \of \thee Green's functions
$G(p_{i},M,\alpha_{0},\alpha_{1},...,\alpha_{n})$ of \thee original
system under renormalization group implies \thee Callan-Symanzik
equations
\begin{eqnarray}
\left(M\frac{\partial}{\partial
M}+\sum_{j=0}^{n}\beta_{j}\frac{\partial}{\partial
\lambda_{j}}+\gamma
\right)G(p_{i},M,\alpha_{0},\alpha_{1},...,\alpha_{n})=0\nonumber
\end{eqnarray}
where $M,\beta_{j},\gamma$ are \thee renormalization mass, \thee
beta functions \andd \thee anomalous dimensions correspondingly.
Similarly \for \thee Green's functions
$G'(p_{i},M,\alpha_{0},\alpha_{1}(\alpha_{0}),...,\alpha_{n}(\alpha_{0}))$
\of \thee reduced system we have
\begin{eqnarray}
\left(M\frac{\partial}{\partial M}+\beta'\frac{\partial}{\partial
\alpha_{0}}+\gamma'
\right)G'(p_{i},M,\alpha_{0},\alpha_{1}(\alpha_{0}),...,\alpha_{n}(\alpha_{0}))=0.\nonumber
\end{eqnarray}
We can see that G$'$ is obtained from G by substituting \thee functions (2.1)
\begin{eqnarray}
G'=G(\alpha_{0},\alpha_{1}(\alpha_{0}),...,\alpha_{n}(\alpha_{0}))\nonumber
\end{eqnarray}
so differentiating with respect to $\alpha_{0}$ we obtain
\begin{eqnarray}
\frac{\partial G'}{\partial \alpha_{0}}=\frac{\partial G}{\partial
\alpha_{0}}+\sum_{j=1}^{n}\frac{\partial G}{\partial
\alpha_{j}}\frac{d \alpha_{j}}{d \alpha_{0}}.\nonumber
\end{eqnarray}
\thee above equations as well as linear independence \of \thee
Green's functions \andd their derivatives lead to \thee relations
\begin{eqnarray}
\beta'= \beta_{0},\qquad \gamma'=\gamma, \qquad \beta' \frac{d \alpha_{j}}{d
\alpha_{0}}=\beta_{j}.\nonumber
\end{eqnarray}
Hence \thee functions (2.1) must satisfy \thee following differential
equations, \thee \reduction equations
\begin{eqnarray}
\beta_{j}= \beta_{0}\frac{d \alpha_{j}}{d \alpha_{0}}.
\end{eqnarray}
A crucial point here is that \thee system (2.2) forms a necessary \andd sufficient condition \for reducing \thee original system \by \thee functions $\alpha_{j}(\alpha_{0})$.

\section{Reduction \of a system with two coupling constants}
For
simplicity we assume that \thee original system has two coupling
\parameters, $\alpha_{0}$ \andd $\alpha_{1}$. We will examine if we can reduce $\alpha_{1}$ in favor \of $\alpha_{0}$. The corresponding beta-functions can
be written \at lowest order as
\begin{eqnarray}
\beta_{0}=b_{0}\alpha_{0}^{2}+....\nonumber
\end{eqnarray}
\begin{eqnarray}
\beta_{1}=c_{1}\alpha_{1}^{2}+
c_{2}\alpha_{0}\alpha_{1}+c_{3}\alpha_{0}^{2}....\nonumber
\end{eqnarray}
which cover a wide range \of models. The \reduction equation which
we have to solve is
\begin{eqnarray}
\beta_{1}= \beta_{0}\frac{d \alpha_{1}}{d \alpha_{0}}.
\end{eqnarray}
Assuming power series solution we can expand $\alpha_{1}$ as
\begin{eqnarray}
\alpha_{1}=p_{0}^{(1)}\alpha_{0}+\sum
_{n=1}p_{n}^{(1)}\alpha_{0}^{(n+1)}.\nonumber
\end{eqnarray}
Substituting \thee above expression into \thee (3.1)
\at lowest order we end up with a quadratic equation
\begin{eqnarray}
c_{1}p_{0}^{2}+(c_{2}-b_{0})p_{0}+c_{3}=0\nonumber
\end{eqnarray}
which can be easily solved calculating \thee corresponding determinant.

\section{Application to \thee MSSM}
\subsection{Relating $\alpha_1$ \andd $\alpha_2$ gauge \couplings}

We will explore \thee possibility to reduce $\alpha_2$ gauge
coupling in favor \of $\alpha_1$  ($\alpha_i=g_i^2/4\pi$). Assuming
that there is a relation between them, a function
$\alpha_2(\alpha_1)$, we have to solve \thee following \reduction
equation
\begin{eqnarray}
\beta_{2}= \beta_{1}\frac{d \alpha_{2}}{d \alpha_{1}}
\end{eqnarray}
where
\begin{eqnarray}
\beta_{2}\equiv \frac{d\alpha_{2}}{dt}=\frac{b_{2}}{2\pi}\alpha_{2}^{2},    \qquad \beta_{1}\equiv \frac{d\alpha_{1}}{dt}=
\frac{b_{1}}{2\pi}\alpha_{1}^{2} \nonumber
\end{eqnarray}
are \thee $\beta$ functions \for \thee $\alpha_2$ \andd $\alpha_1$
gauge \couplings correspondingly, $b_2=1$ \andd $b_1=11$ are \thee
$\beta$ function coefficients \andd $t=\ln E$. Assuming that \thee
differential equation (4.1) has a power series solution, we can
expand $\alpha_{2}$ \at lowest order in perturbation theory as
\begin{eqnarray}
 \alpha_{2}=c_{0}\alpha_{1}       \nonumber
\end{eqnarray}
where $c_0$ is a constant.
Substituting this relation to \thee reduction equation (4.1) we are led to
\begin{eqnarray}
 c_{0}=\frac{\beta_{2}}{\beta_{1}}=\frac{b_{2}\alpha_{2}^{2}}{b_{1}\alpha_{1}^{2}}= \frac{b_{2}c_{0}^{2}\alpha_{1}^{2}}{b_{1}\alpha_{1}^{2}} \Rightarrow   \nonumber
\end{eqnarray}
\begin{eqnarray}
 c_{0}(c_{0}b_{2}- b_{1})=0 \Rightarrow\nonumber
\end{eqnarray}
\begin{eqnarray}
 c_{0}=0, \qquad c_{0}=11.\nonumber
\end{eqnarray}
Hence $\alpha_{2}$ can be written as a function \of $\alpha_{2}$ as
\begin{eqnarray}
 \alpha_{2}=11\alpha_{1}.\nonumber
\end{eqnarray}
We can check now if \thee above result is compatible with \thee experimental
values
\begin{eqnarray}
 \frac{1}{\alpha_{em}(Mz)}=\frac{1}{\alpha_{1}(Mz)}+\frac{1}{\alpha_{2}(Mz)}\Rightarrow \nonumber
\end{eqnarray}
\begin{eqnarray}
\alpha_{em}(Mz)=\frac{11}{12}\alpha_{1}(Mz).\nonumber
\end{eqnarray}
We know that
\begin{eqnarray}
\sin^{2}\theta_{w}(Mz)=\frac{\alpha_{em}(Mz)}{\alpha_{2}(Mz)}\Rightarrow\nonumber
\end{eqnarray}
\begin{eqnarray}
\sin^{2}\theta_{w}(Mz)=\frac{11}{12}\frac{\alpha_{1}(Mz)}{11\alpha_{1}(Mz)}=\frac{1}{12}=0.08333\nonumber
\end{eqnarray}
which is unacceptable because
$\sin^{2}\theta_{w}(Mz)_{exp}=0.23151\pm 0.00017$. Concluding,
\thee \reduction of $\alpha_{2}$ \andd $\alpha_{1}$ \couplings in
\thee context \of \thee MSSM is not possible.

\subsection{Relating $\alpha_t$ top quark \andd $\alpha_b$ bottom quark Yukawa \couplings}

Following \thee same procedure we assume that $\alpha_{t}$
Yukawa coupling can be related with \thee $\alpha_{b}$ Yukawa
coupling ($\alpha_{i}=h_i^2/4\pi,i=t,b$), so they must satisfy \thee \reduction equation
\begin{eqnarray}
\beta_{t}= \beta_{b}\frac{d \alpha_{t}}{d
\alpha_{b}}\Rightarrow\nonumber
\end{eqnarray}
\begin{eqnarray}
\frac{d \alpha_{t}}{d
\alpha_{b}}=\frac{\beta_{t}}{\beta_{b}}=\frac{\alpha_{t}(6\alpha_{t}+\alpha_{b}-\frac{13}{15}\alpha_{1}-3\alpha_{2}+\frac{16}{3}\alpha_{3})}{\alpha_{b}(6\alpha_{b}+\alpha_{t}+\alpha_{\tau}-\frac{7}{15}\alpha_{1}-3\alpha_{2}+\frac{16}{3}\alpha_{3})}\nonumber
\end{eqnarray}
where $\beta_{t}$ \andd $\beta_{b}$ are \thee $\beta$ functions \of
top quark Yukawa coupling \andd bottom quark Yukawa coupling
correspondingly. We can \for simplicity neglect \thee contribution
from \thee $\tau$ lepton, $\alpha_{\tau}$,  \andd \thee small
difference between $\frac{13}{15}$ \andd $\frac{7}{15}$, so we are
led to
\begin{eqnarray}
\frac{\beta_{t}}{\beta_{b}}=\frac{\alpha_{t}(6\alpha_{t}+\alpha_{b}-\frac{13}{15}\alpha_{1}-3\alpha_{2}+\frac{16}{3}\alpha_{3})}{\alpha_{b}(6\alpha_{b}+\alpha_{t}-\frac{13}{15}\alpha_{1}-3\alpha_{2}+\frac{16}{3}\alpha_{3})}
\end{eqnarray}
Assuming again power series solution \of \thee \reduction equation
we can expand top quark Yukawa coupling \at lowest order as
\begin{eqnarray}
\alpha_{t}=d_{0}\alpha_{b}\nonumber
\end{eqnarray}
where $d_{0}$ is a constant. The derivative \of \thee ratio \of
\thee two Yukawa \couplings must be zero
\begin{eqnarray}
   \frac{d}{dt}(\frac{\alpha_{t}}{\alpha_{b}})=0\Rightarrow                                  \nonumber
\end{eqnarray}
\begin{eqnarray}
   \frac{1}{\alpha_{t}^{2}}(\alpha_{b}\beta_{t}-\alpha_{t}\beta_{b})=0\Rightarrow                                  \nonumber
\end{eqnarray}
\begin{eqnarray}
  \frac{\alpha_{t}}{\alpha_{b}}=\frac{\beta_{t}}{\beta_{b}}.                                 \nonumber
\end{eqnarray}
Substituting \thee above result into eqn.(4.2) we obtain
\begin{eqnarray}
\frac{\alpha_{t}}{\alpha_{b}}=\frac{\alpha_{t}(6\alpha_{t}+\alpha_{b}-\frac{13}{15}\alpha_{1}-3\alpha_{2}+\frac{16}{3}\alpha_{3})}{\alpha_{b}(6\alpha_{b}+\alpha_{t}-\frac{13}{15}\alpha_{1}-3\alpha_{2}+\frac{16}{3}\alpha_{3})}\Rightarrow
\nonumber
\end{eqnarray}
\begin{eqnarray}
6\alpha_{t}+\alpha_{b}-\frac{13}{15}\alpha_{1}-3\alpha_{2}+\frac{16}{3}\alpha_{3}=6\alpha_{b}+\alpha_{t}-\frac{13}{15}\alpha_{1}-3\alpha_{2}+\frac{16}{3}\alpha_{3}\Rightarrow
 \nonumber
\end{eqnarray}
\begin{eqnarray}
\alpha_{t}=\alpha_{b}.
 \nonumber
\end{eqnarray}
That is, if we start with  $\alpha_{t}$ \andd $\alpha_{b}$ equal \at an energy
scale, equality will persist
 \at all energies.

The next thing to do is to solve numerically \thee one-loop coupled
differential equations \of top \andd bottom Yukawa \couplings taken
account \thee $\alpha_{\tau}$ contribution \andd \thee difference
between \thee numerical factors, to see
 if such a relation like \thee previous one can exist.
 First, we solve \thee differential equations
\for \thee gauge \andd Yukawa couplings in \thee SM. And then at
$M_{SUSY}$ we impose \thee next boundary conditions \for some values
\of $\tan\beta$ that keeps \thee ratio $\alpha_{t}/\alpha_{b}$
constant \for all energies
\begin{eqnarray}
\alpha_{t_{SM}}(M_{SUSY})=\alpha_{t_{MSSM}}(M_{SUSY})\sin^{2}\beta
 \nonumber
\end{eqnarray}
\begin{eqnarray}
\alpha_{b_{SM}}(M_{SUSY})=\alpha_{b_{MSSM}}(M_{SUSY})\cos^{2}\beta
 \nonumber
\end{eqnarray}
\begin{eqnarray}
\alpha_{\tau_{SM}}(M_{SUSY})=\alpha_{\tau_{MSSM}}(M_{SUSY})\cos^{2}\beta
 \nonumber
\end{eqnarray}

\begin{figure}[!t]
\begin{tabular}{cc}
\includegraphics[scale=0.45,angle=0]{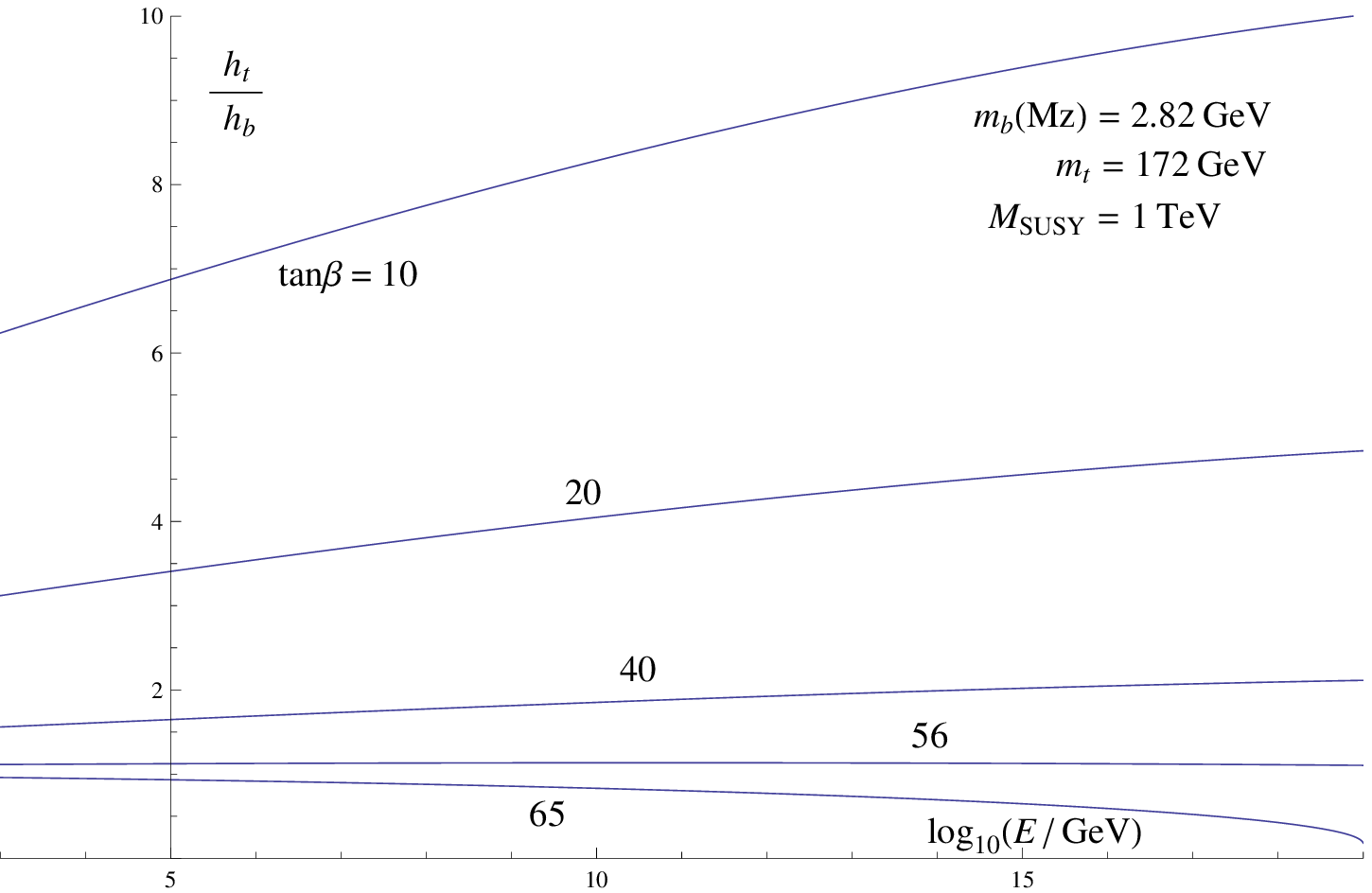}&
\includegraphics[scale=0.45,angle=0]{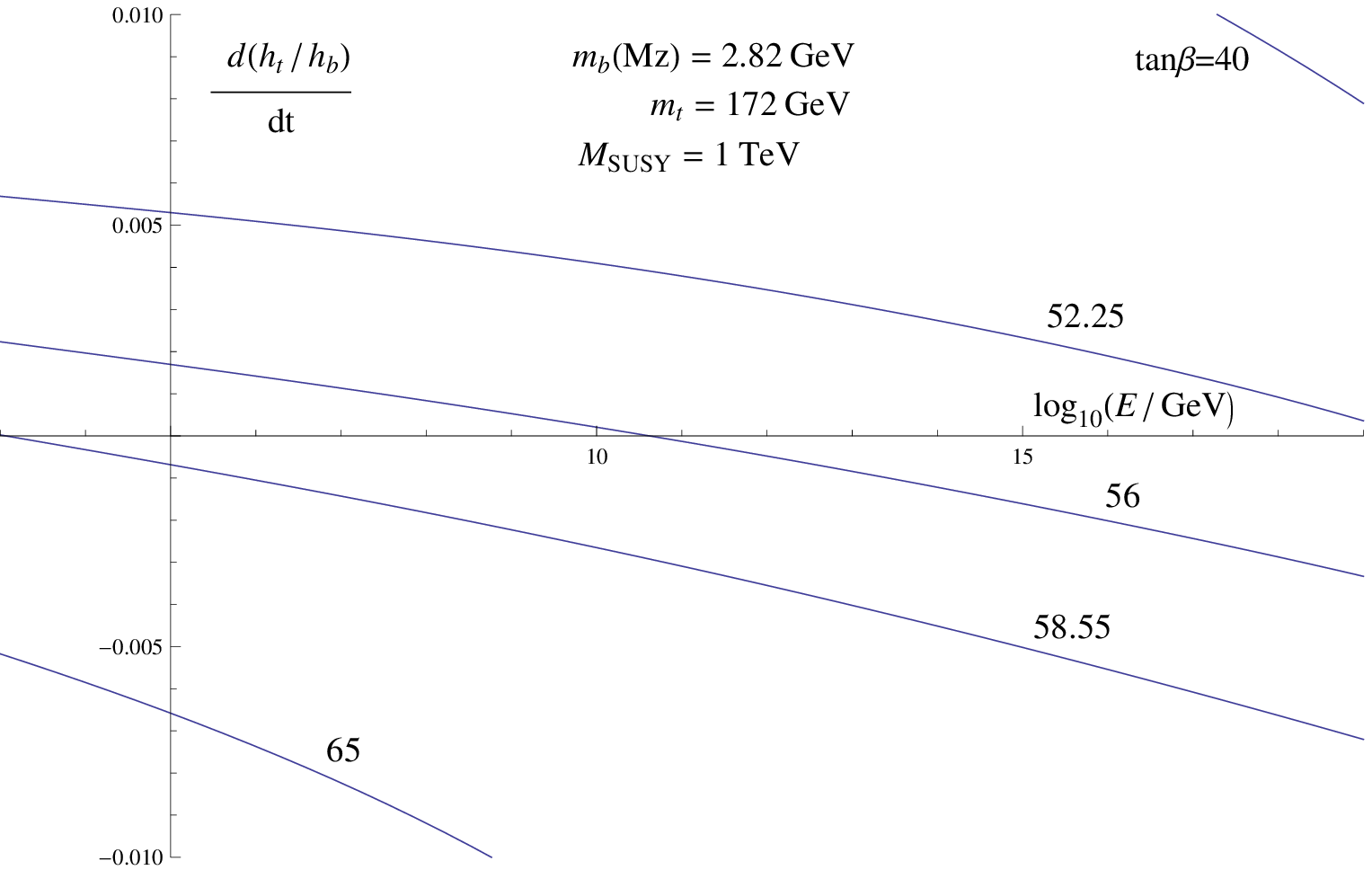}\\
(a)&(b)
\end{tabular}
\caption{(a) The ratio \of top \andd bottom quark Yukawa \couplings
$h_t/h_b$ \andd (b) \thee derivative \of their ratio as a function
\of energy. Several values \of $\tan\beta$ have been taken in
addition with $M_{SUSY}=1$~TeV, $m_t=172$ GeV \andd $m_b(M_Z)=2.82$
GeV.}
\label{fig:htoverhb_1}
\end{figure}

In Fig.\ref{fig:htoverhb_1} we plot \thee ratio \of \thee two Yukawa
\couplings under investigation $h_t/h_b$ (a) \andd \thee derivative
\of their ratio (b) as a function of energy, \for several values \of
$\tan\beta$, $M_{SUSY}=1$~TeV, $m_b(M_Z)=2.82$ GeV \andd $m_t=172$
GeV. We can see that \for \thee range \of values \for $\tan\beta$
between
52.25-58.55, \thee derivative \of \thee ratio is very close to zero. 

\begin{figure}[!h]
\begin{tabular}{cc}
\includegraphics[scale=0.47,angle=0]{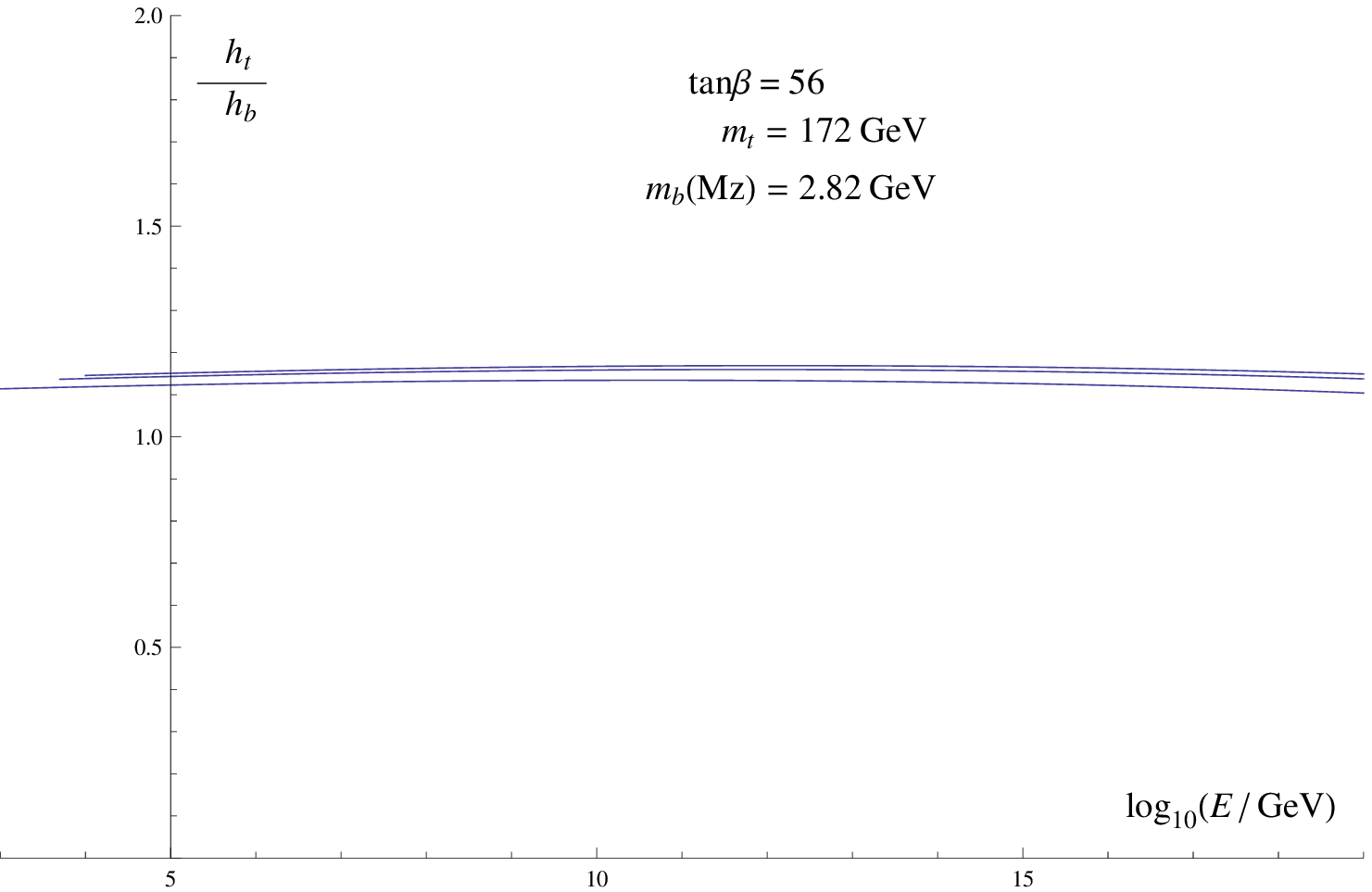}&
\includegraphics[scale=0.47,angle=0]{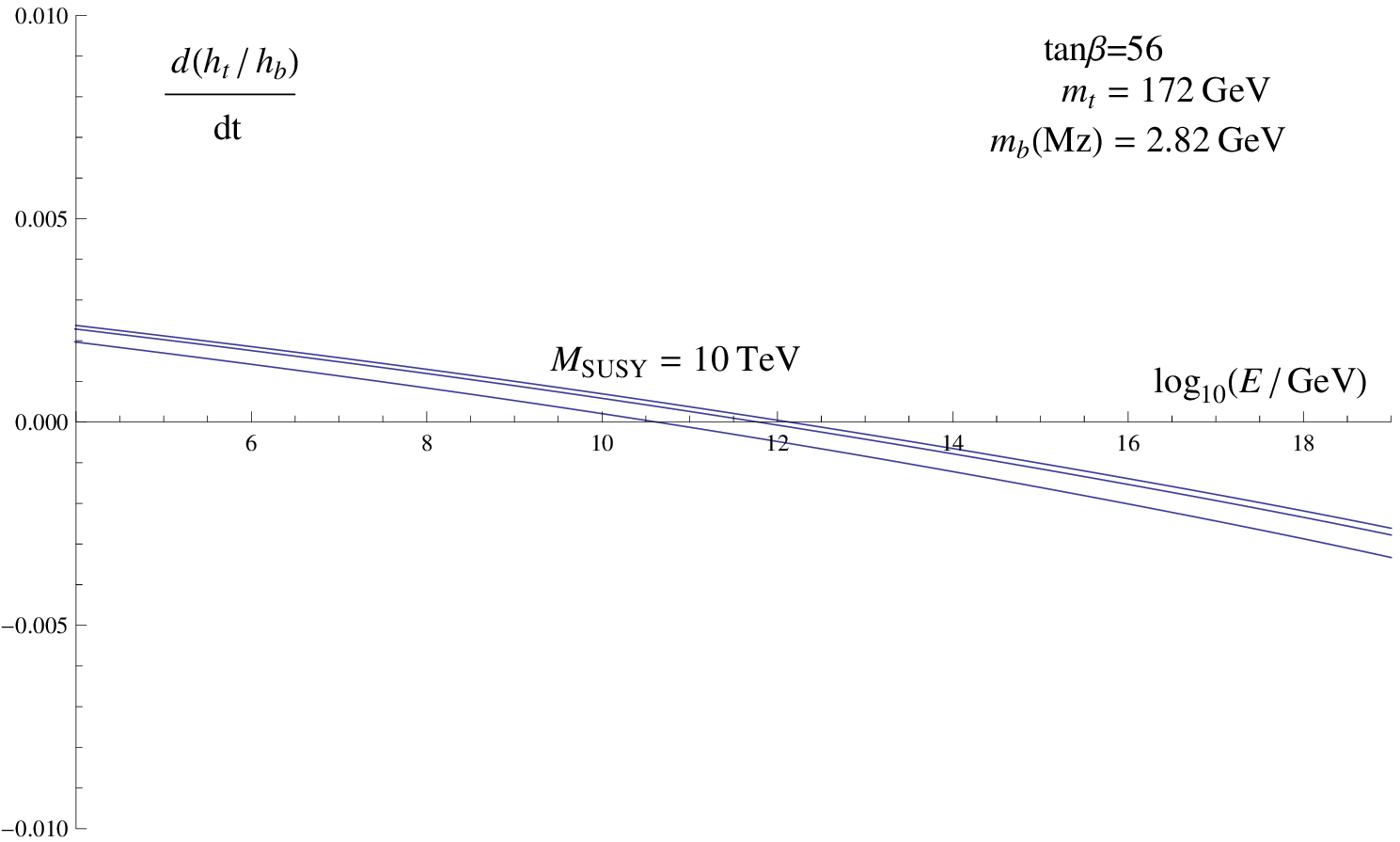}\\
(a)&(b)\\
\includegraphics[scale=0.47,angle=0]{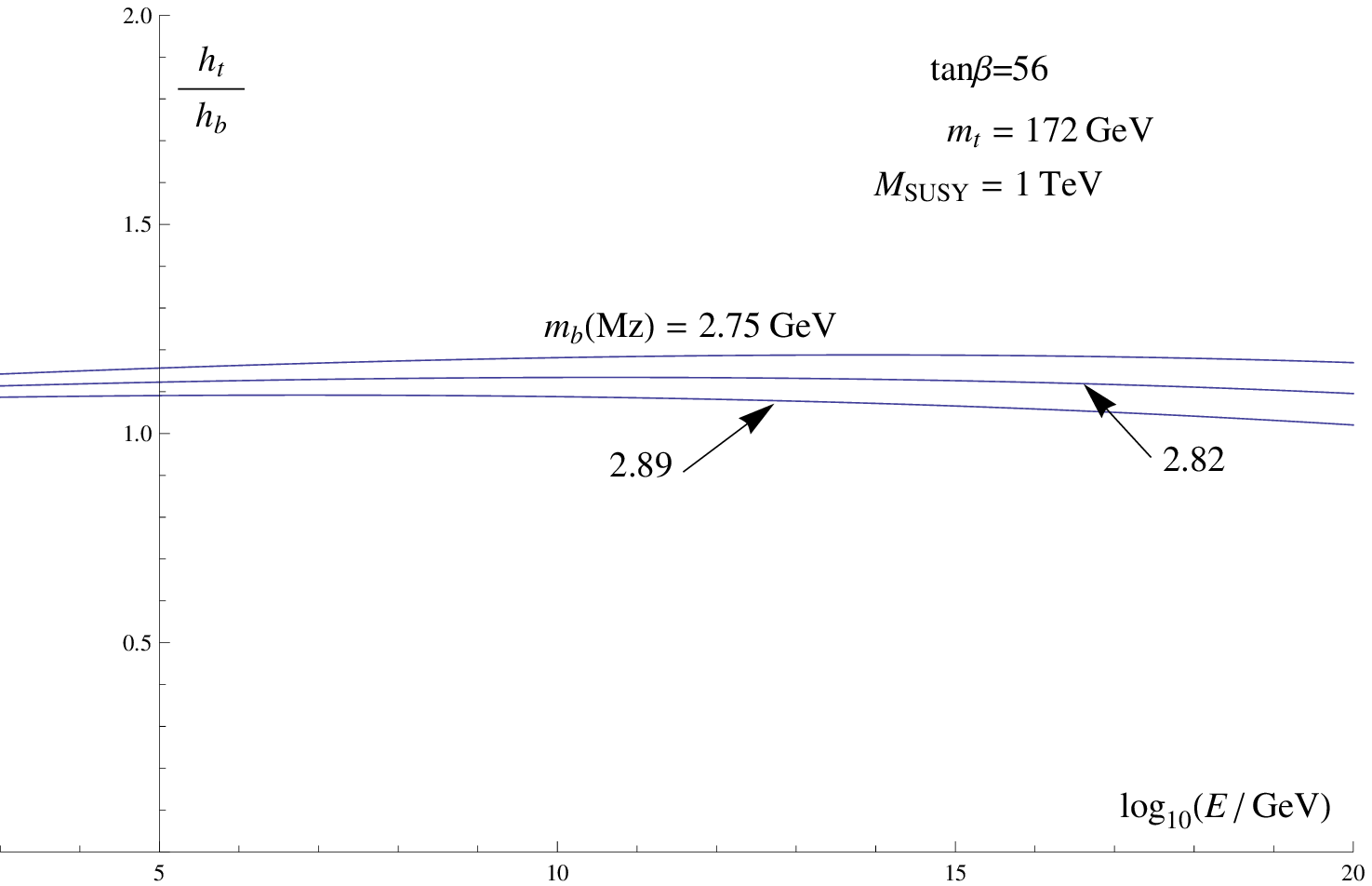}&
\includegraphics[scale=0.47,angle=0]{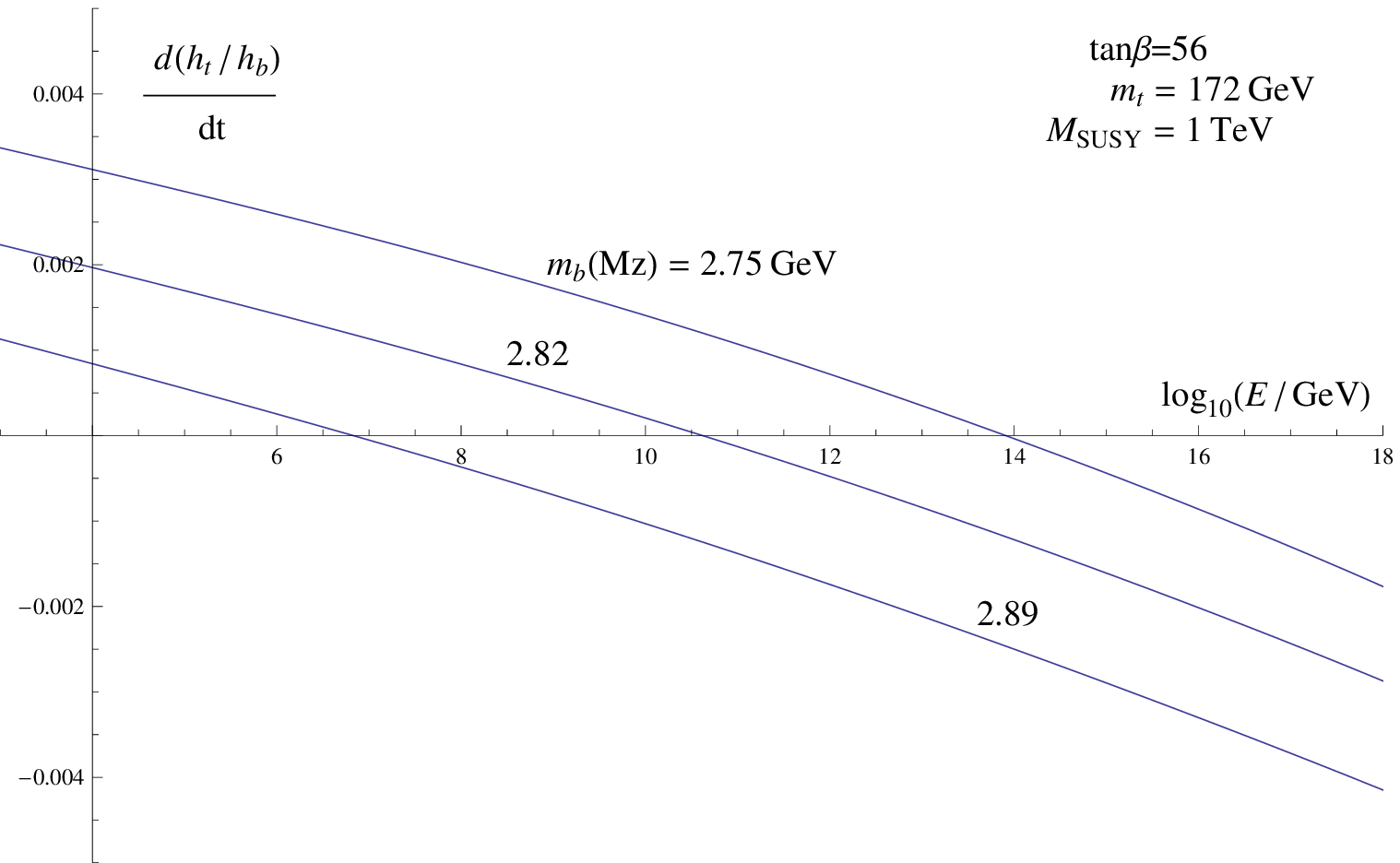}\\
(c)&(d)
\end{tabular}
\caption{(a) The ratio \of top \andd bottom quark Yukawa \couplings
$h_t/h_b$ \andd (b) \thee derivative \of their ratio as a function
\of energy \for $M_{SUSY}=1$, 5 \andd 10 TeV. (c) The ratio
$h_t/h_b$ \andd (d) \thee derivative \of \thee ratio as a function
of energy \for $M_{SUSY}=1$ TeV \andd three values \of \thee bottom
mass that vary in \thee experimental error region.}
\label{fig:htoverhb_2}
\end{figure}

In Fig.\ref{fig:htoverhb_2} in (a) \andd (c) we plot \thee ratio
$h_t/h_b$  as well as \thee
 derivative \of \thee ratio in (b) \andd (d) as a function \of  energy \for $\tan\beta=56$. In (a) \andd
(b) we can see three curves corresponding to $M_{SUSY}=1$, 5 \andd
10 TeV (we have kept \thee masses \of top \andd bottom quarks at
their central values). In (c) \andd (d) we have taken three values
\for \thee bottom mass $m_b(M_Z)=2.75$, 2.82 \andd 2.89 GeV, keeping
\thee top mass \at its central value \andd $M_{SUSY}=1$ TeV.
\newpage

\newpage

\begin{figure}[!h]
\begin{tabular}{cc}
\includegraphics[scale=0.55,angle=0]{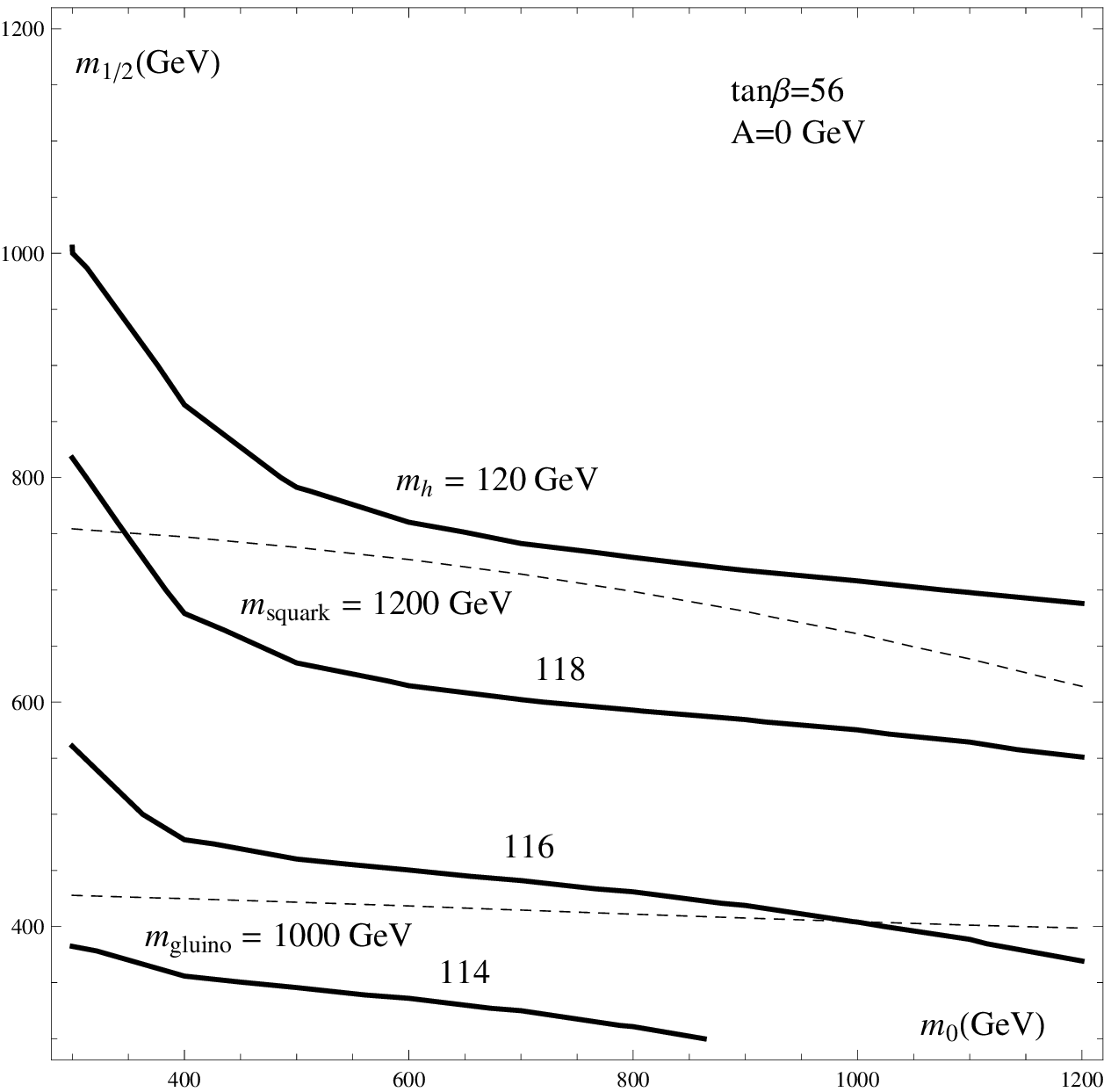}&
\includegraphics[scale=0.55,angle=0]{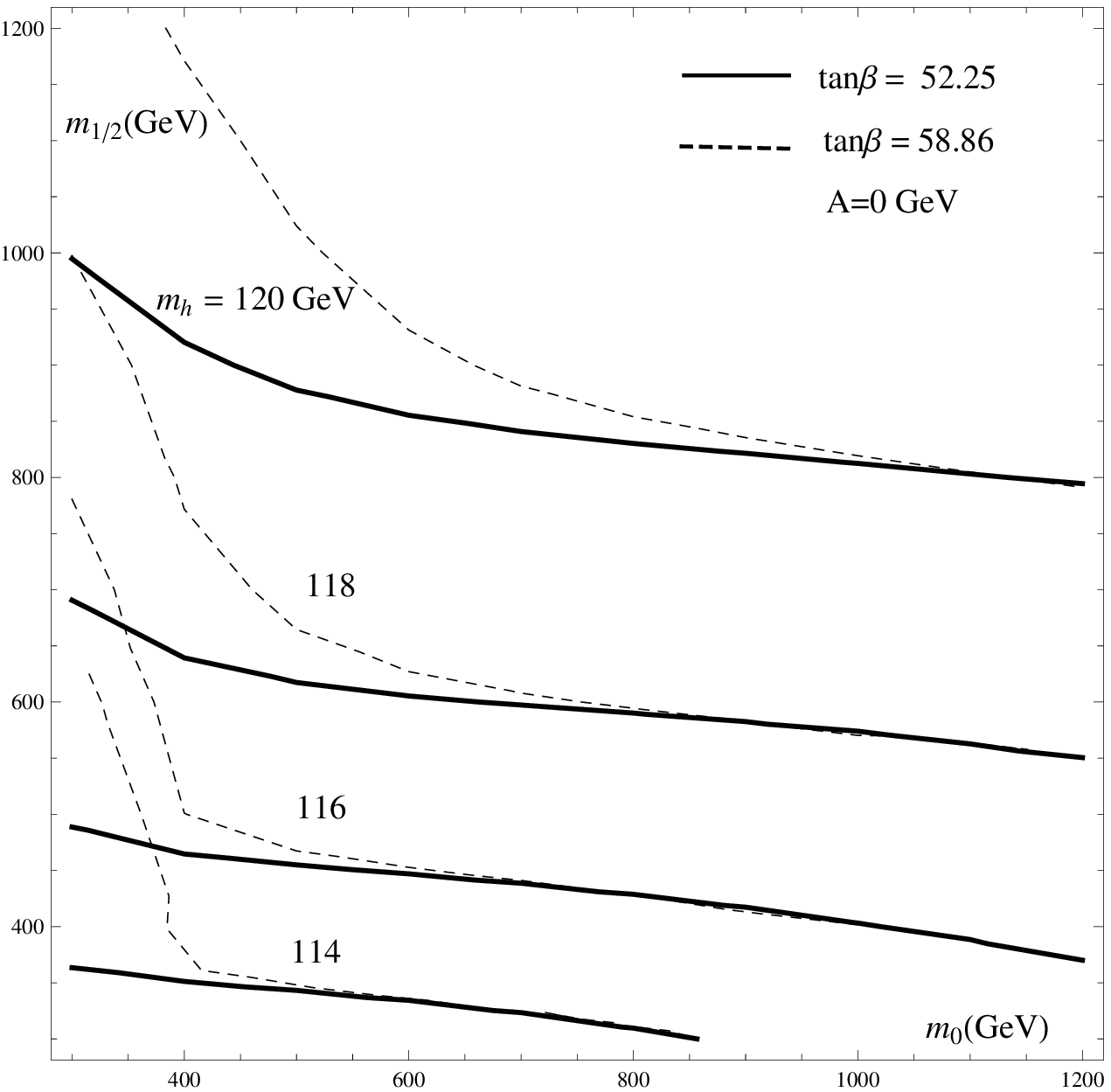}\\
(a)&(b)
\end{tabular}
\caption{(a) Contours \of constant pole mass \for \thee lightest
supersymmetric Higgs $m_h$ in \thee plane of $(m_0,m_{1/2})$ \for
initial value $A=0$ GeV \andd \for $\tan\beta=56$. The contours with
\thee dashed line correspond to a gluino mass \of 1 TeV \andd to
(\thee lightest) squark mass of 1.2 TeV correspondingly. In (b) we
plot contours \of constant $m_h$ (pole) mass in \thee plane \of
$(m_0,m_{1/2})$ \for initial value $A=0$ GeV \andd \for two values
\of $\tan\beta$: 58.55 \andd 52.25.}
\label{fig:htoverhb_123}
\end{figure}

Having used a Fortran code \for \thee calculation of higgs particle spectrum in \thee MSSM (SUSPECT \cite{SUSPECT})\footnote{%
We run \thee program choosing \thee mSUGRA model, running \thee
renormalization group equations in 2-loop level \andd evaluating
\thee pole masses. Also we choose $\textrm{sign}(\mu)=+1$.}, we plot
in \thee plane \of sfermions \andd gauginos, $(m_0,m_{1/2})$,
contours \of constant mass values \for \thee lightest supersymmetric
Higgs $m_h$. In Fig.\ref{fig:htoverhb_123} in (a) we plot these
contours \for several values \of \thee lightest supersymmetric Higgs
$m_h=114,116,118,120$ GeV. We also choose \thee values $A=0$ GeV
\andd $\tan\beta=56$. The contours with \thee dashed line correspond
to a gluino mass \of 1 TeV \andd to (\thee lightest) squark mass \of
1.2 TeV correspondingly. In (b) we plot contours \of constant mass
values \for \thee lightest supersymmetric Higgs $m_h$ \for two
values of $\tan\beta$: 58.55 \andd 52.25.

\section{Conclusions}

The program \of \reduction \of \couplings is a very powerful method
that relates arbitrary coupling \parameters. As a result we obtain a
new reduced theory which has an increased predictive power. This
method has been applied to supersymmetric GUTs \andd lately to \thee
MSSM with great success. In our work relating top \andd bottom quark
Yukawa \couplings, we give a narrow range \of values \for
$\tan\beta$ that can permit us to give a prediction \for \thee
lightest supersymmetric Higgs particle in \thee MSSM.

\newpage

\noindent \textbf {Acknowledgments}
\\

This talk is based on a work that has been done in collaboration
with N.D. Tracas, N.D. Vlachos \andd G. Zoupanos. G.T. would like to
thank them \for valuable discussions. G.T. is grateful \for the kind
hospitality \of \thee organizers \of \thee Summer School \andd
Workshop on \thee Standard Model \andd Beyond, held in Corfu.



\end{document}